\documentclass[compsoc, conference, a4paper, 10pt, times]{IEEEtran}

\usepackage{cite}
\usepackage{amsmath,amssymb,amsfonts}
\usepackage{algorithmic}
\usepackage{graphicx}
\usepackage{textcomp}
\usepackage{bmpsize}
\usepackage[frozencache]{minted}
\usepackage[colorlinks=true, urlcolor=black]{hyperref}

\usepackage[dvipsnames]{xcolor}
\usepackage{tikz}
\usetikzlibrary{shapes.multipart, positioning, calc, decorations.pathreplacing, fit}

\tikzstyle{block} = [
    text centered,
    draw=black,
    minimum height=1cm,
    text width=3cm,
    node distance=0.5cm and 3cm
]

\tikzstyle{party} = [
    block,
    rectangle split,
    rectangle split parts=2,
    rounded corners,
    thick
]

\tikzstyle{pstep} = [
    block,
]

\begin{document}

\title{Reverse Online Guessing Attacks on PAKE Protocols}

\author{%
  \IEEEauthorblockN{%
    Eloise Christian \\
    Tejas Gadwalkar \\
    Arthur Azevedo de Amorim}
  \IEEEauthorblockA{%
    Rochester Institute of Technology
  }
  \and
  \IEEEauthorblockN{Edward V. Zieglar Jr.}
  \IEEEauthorblockA{National Security Agency}
}

\maketitle

\begin{abstract}
Though not yet widely deployed, \emph{password-authenticated key exchange} (PAKE) protocols have
been the subject of several recent standardization efforts, partly because of their resistance
against various guessing attacks,
but also because they do not require a public-key infrastructure (PKI), making them naturally
resistant against PKI failures.

The goal of this paper is to reevaluate the PAKE model by noting that the absence of a PKI---or,
more generally, of a mechanism aside from the password for authenticating the server---makes such
protocols vulnerable to \emph{reverse online guessing attacks}, in which an adversary attempts to
validate password guesses by impersonating a server. While their logic is similar to traditional
guessing, where the attacker impersonates a client, reverse guessing poses a unique risk because the
burden of detection is shifted to the clients, rendering existing defenses against traditional
guessing moot. Our results demonstrate that reverse guessing is particularly effective when an
adversary attacks clients indiscriminately, such as in phishing or password-spraying attacks, or for
applications with automated login processes or a universal password, such as WPA3-SAE. 
Our analysis suggests that stakeholders should, by default, authenticate the server using more
stringent measures than just the user's password, and that a password-only mode of operation should
be a last resort against catastrophic security failures when other authentication mechanisms are not
available.
\end{abstract}


\section{Introduction}
\label{sec:introduction}
Despite its many flaws, password-based authentication remains the primary mechanism for protecting
user accounts. In a typical web application, the user logs in by establishing an encrypted,
authenticated TLS channel with the application server and sending their password. The server
combines the password with a unique salt value associated with the user's account, hashes the result
and ensures that the hashed password matches what is stored in its password file. This method works
pretty well, but it has several important drawbacks. First, during login, the server has access to
the password in plain text, which could lead to leaks in case of programming errors. Second, the
method is susceptible to an offline dictionary attack if the server's password file is leaked. An
attacker can combine the leaked salt values with a dictionary of common passwords to find candidate
hashed passwords for each user. If a candidate matches what is stored in the password file, the
attacker will have learned the user password. Finally, the password can be leaked due to public-key
infrastructure (PKI) failures, such as if the client checks the server's certificate incorrectly, or
if the server's private key is stolen.

These limitations have led to the development of \emph{password-authenticated key exchange} (PAKE)
protocols, where the password is only seen by the user and is not directly exposed to the server or
other parties. PAKE protocols are usually \emph{password-only}: they authenticate both the client
and server using only the password, eliding PKI entirely.
For example, the Encrypted Key Exchange (EKE) protocol \cite{eke} uses a combination of symmetric
and asymmetric encryption to prevent eavesdropped handshake messages from revealing anything about
the underlying password; cracking a password is only possible by carrying brute-force guesses
against random encryption keys, which is much harder.

Of course, because passwords have low entropy, compromise remains possible by way of \emph{guessing
attacks}. In an \emph{online} attack, the adversary impersonates a client and launches successive
login attempts with different password guesses until getting the guess right. This attack is
inevitable since the adversary uses the same basic means to attempt authentication as an honest user
would.
To counteract this attack, there are a variety of server-side techniques which take advantage of the
fact that the server can see and track all of the adversary's failed login attempts and dynamically
respond to them. For instance, it could block further requests from certain IP addresses,
temporarily lock targeted accounts, or enforce Automated Turing Tests (e.g. CAPTCHAs) \cite{pgrp,
StopGuessing, arana, nist-dig}. While these are by no means perfect (e.g. blocking accounts
trivially leads to a denial of service attack), developing these techniques is an active area of
research in the literature.



The goal of this paper is to draw attention to another, more insidious threat to password-only PAKE
protocols: what we call \emph{reverse} online guessing attacks. If the client's only mechanism to
authenticate the server is data derived from the password itself, an adversary can impersonate the
server instead of the client, intercept client authentication requests, and complete handshakes
using password guesses.
In contrast to the standard attack, there is little mention of the reverse attack in the literature,
let alone a detailed analysis of its practical risks. Superficially, it may even seem that the two
forms of guessing are similar and that reverse guessing hardly requires special attention.

However, our main contribution is to show that, despite the surface-level similarities and the
simple concept behind reverse guessing, this attack poses a distinct threat to password-only
protocols. In particular, the server-side techniques for combating the standard attack do not apply
to the reverse attack as the server is uninvolved in the attack flow.
As we will see, integrating stringent server authentication prevents reverse online guessing, and
PKI is one of the simplest ways to do this; however, practitioners may overlook the importance of
this measure precisely because many of these protocols are advertised as PKI-free \cite{aeke,
pake-sok, opaque}, and because the risks of reverse guessing have not been properly exposed.
Even in settings where no trusted PKI can be integrated, we expect our work to serve as a guideline
for current and future PAKE standardization efforts, as well as to reevaluate the security of PAKE
protocols in practice (cf. \autoref{sec:practical-attacks}) and reconsider the importance of
explicit server authentication.

\subsection{Threat Model}
\label{sec:threat-model}
Given that protocols can be deployed in a variety of unpredictable settings, we argue that the
behavior of attackers should not be assumed to be restricted in ad hoc ways.
%
%
Following the Dolev-Yao model \cite{dy}, we assume a powerful attacker that can intercept several
login requests from users via social engineering or compromising large amounts of network
infrastructure. 
We assume that the attacker is not interested in targeting specific users, but rather a large group
indiscriminately. Possible actors which could fit this model include nation-states or other wealthy
organizations that are interested in controlling the online identities of several users in a region.
While this limits the applicability of the vulnerabilities we will discuss, there have been several
high-profile attacks exposed that fall under these characteristics. One such example is the recent
SolarWinds supply-chain compromise which affected government agencies and businesses around the
world \cite{solarwinds-perspectives}.


\subsection{Contributions and Structure of the Paper}
In sum, our contributions are:

\begin{enumerate}
    \item The first comprehensive description of reverse online guessing attacks against
    password-only PAKE protocols (\autoref{sec:attack}). For concreteness, we present the attack in
    the context of the EKE protocol~\cite{eke}, while evincing a general structure that can be
    applied to several other protocols of theoretical and practical interest.

    \item An analysis of the key advantages that make this attack a particularly severe threat
    compared to traditional online guessing (\autoref{sec:comparison}), as well as of particularly
    risky scenarios where this attack might prove useful (\autoref{sec:practical-attacks}).

    \item A discussion of possible mitigation techniques against reverse online guessing and their
    shortcomings (\autoref{sec:mitigation}).

    \item A technique for identifying the presence of this attack in several protocols using
    automated checkers in the symbolic model of cryptography. We have applied our technique to
    expose this vulnerability in EKE \cite{eke}, A-EKE \cite{aeke}, SRP versions 3 and 6a
    \cite{srp3, srp6a-rfc}, Dragonfly Key Exchange \cite{dragonfly}, and OPAQUE \cite{opaque}.
    The technique also demonstrates that variants of the protocols which include server
    authentication (e.g. SRP-6a in some modes defined in RFC 5054) are immune to the attack
    (\autoref{sec:symbolic-analysis}).
\end{enumerate}

We conclude in \autoref{sec:conclusion}.

\subsection{Related Work}
\label{sec:related-work}
We found few references that acknowledge the possibility of reverse online guessing. First, a
working draft of the Internet Research Task Force (IRTF) to standardize the OPAQUE protocol notes
the inevitable possibility of online guessing attacks orchestrated by malicious clients or servers
\cite{opaque-irtf}. The draft explains that, in both settings, the attacks are exhaustive and
equally expensive, and furthermore, that attacks from the server side can be mitigated with server
authentication, such as via TLS.
Another analysis of OPAQUE concludes similarly but notes that server authentication ``\emph{would
not be in line with the `spirit' of OPAQUE}'' (emphasis theirs) \cite{whatsapp}. Indeed, the
protocol's intention is that ``the security of passwords and password authentication does not rely
on PKI but on OPAQUE only'' \cite{opaque}.
Finally, a critique of OPAQUE's publication \cite{opaque} and standardization draft
\cite{opaque-irtf} notes the attack is possible and does not require access to the server's private
data \cite{opaque-critique}.

Separately, a textbook on key exchange protocols explains the attack's threat to a three-party PAKE
protocol, unless client-side mitigation is employed, which is unusual. It adds that clients which
cache the password and retry logins may exacerbate the issue \cite{boyd2020protocols}[\S 8]. We
expand on these points in \autoref{sec:comparison} and \autoref{sec:attack-wifi}.

Our work is consistent with these analyses, but also highlights the practical risks that could arise
in the absence of server authentication.

\section{Attack Overview}
\label{sec:attack}
Reverse online guessing attacks exploit the structure of PAKE protocols to turn clients into
password oracles. The root cause of the vulnerability is that the adversary can forge all the data
required to run a handshake based on a password guess. The specific messages or computations
involved in the protocols do not play an important role. For concreteness, we present the general
idea in detail using Encrypted Key Exchange (EKE) \cite{eke}, as it is a particularly simple PAKE
protocol. Later, we discuss how the attack can be adapted to other popular PAKE protocols.

\subsection{Encrypted Key Exchange}
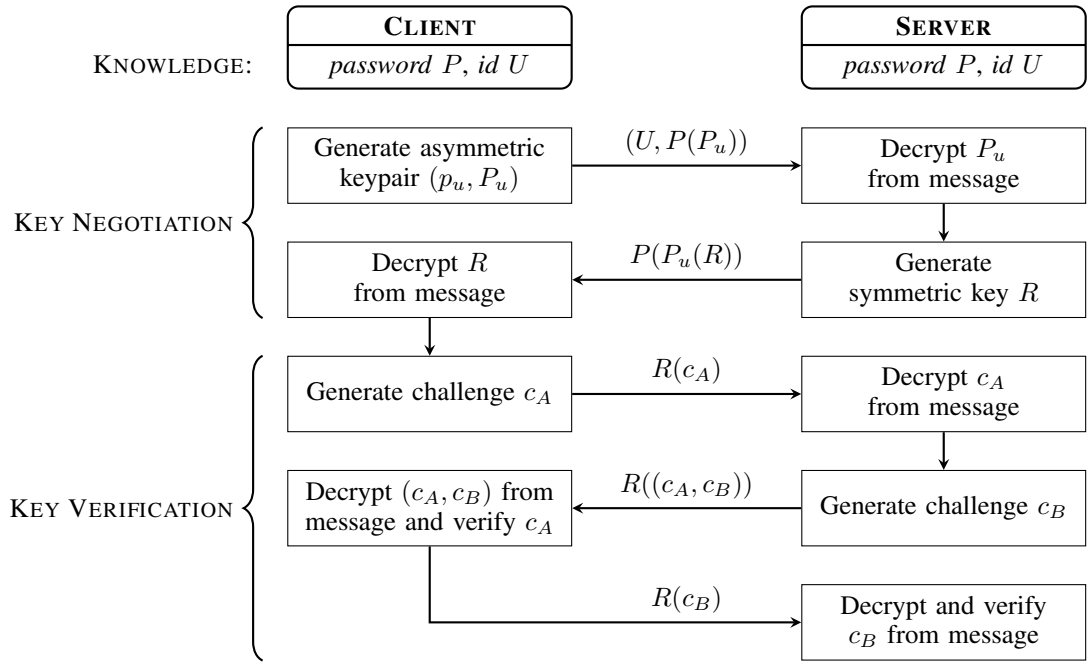
\begin{figure*}[!t]
    \centering
    \begin{tikzpicture}[
    block/.append style={text width=3.5cm},
    phase/.style={thick, decorate, decoration={brace, mirror, amplitude=0.25cm}},
    phaselabel/.style={midway, left, xshift=-0.25cm, font=\scshape},
]
    \node (client) [party]
    {\textbf{\textsc{Client}} \nodepart{two} \textit{password} $P$, \textit{id} $U$};
    \node (server) [party, right=of client]
    {\textbf{\textsc{Server}} \nodepart{two} \textit{password} $P$, \textit{id} $U$};

    \node (plabel) [font=\scshape]
    at ($(client.north west)!0.7!(client.south west) - (1.5cm, 0)$)
    {Knowledge:};

    \node (init) [pstep, below=of client] {Generate asymmetric keypair $(p_u, P_u)$};

    \node (proc) [pstep, below=of server] {Decrypt $P_u$ from message};
    \node (send) [pstep, below=of proc] {Generate symmetric key $R$};

    \node (recv) [pstep] at (init |- send) {Decrypt $R$ from message};

    \node (cprompt) [pstep, below=of recv] {Generate challenge $c_A$};

    \node (ssolve) [pstep, below=of send] {Decrypt $c_A$ from message};
    \node (sprompt) [pstep, below=of ssolve] {Generate challenge $c_B$};

    \node (csolve) [pstep, below=of cprompt] {Decrypt $(c_A, c_B)$ from message and verify $c_A$};

    \node (sverify) [pstep, below=of sprompt] {Decrypt and verify $c_B$ from message};

    \draw[thick, -stealth] (init) -- (proc) node [midway, above] {$(U, P(P_u))$};
    \draw[thick, -stealth] (proc) -- (send);
    \draw[thick, -stealth] (send) -- (recv) node [midway, above] {$P(P_u(R))$};
    \draw[thick, -stealth] (recv) -- (cprompt);
    \draw[thick, -stealth] (cprompt) -- (ssolve) node [midway, above] {$R(c_A)$};
    \draw[thick, -stealth] (ssolve) -- (sprompt);
    \draw[thick, -stealth] (sprompt) -- (csolve) node (smsg) [midway, above] {$R((c_A, c_B))$};
    \draw[thick, -stealth] (csolve) |- (sverify) node [above] at (smsg |- sverify) {$R(c_B)$};

    \draw[phase] (plabel.east |- init.north west) -- (plabel.east |- recv.south west)
    node [phaselabel] {Key Negotiation};
    \draw[phase] (plabel.east |- cprompt.north west) -- (plabel.east |- sverify.south west)
    node [phaselabel] {Key Verification};
\end{tikzpicture}
    \caption{
        The Encrypted Key Exchange \cite{eke} protocol. A client and server, with knowledge of a
        shared password $P$, negotiate and verify a session key $R$.
    }
    \label{fig:eke}
\end{figure*}

\autoref{fig:eke} describes the flow of Encrypted Key Exchange \cite{eke}. The protocol assumes that
the client and server already know the client's identifier $U$ and the shared password $P$. In other
protocols, the server does not have direct access to the password, but only to derived information
which allows the client to authenticate, such as a hash of $P$ and some salt $s$ \cite{pake-sok}. We
call this \emph{authentication data}, or \emph{auth data} for short. In practice, this data is
exchanged or generated during an initial \emph{registration phase}, but the details of this phase
are unimportant in what follows. Key exchange proceeds in two phases: \emph{key negotiation} and
\emph{key verification}.

During key negotiation, the parties exchange messages to establish the session key which will secure
their communications. The client initiates this phase by sending a \textit{request} to the server.
In EKE, this request consists of the client's identifier $U$, in cleartext, and a fresh public key
$P_u$, encrypted with $P$. The server processes this request and replies with an appropriate
\textit{response}. In EKE, this response consists of a fresh symmetric key $R$, doubly encrypted
with $P_u$ and $P$. The client processes this response to find the session key.  

For security reasons, negotiation is often not enough for a party to check the validity of the
password. In EKE, for example, it is not possible for the server to tell whether the decryption of
$P(P_u)$ succeeds or not; otherwise, the protocol would be vulnerable to an offline guessing attack,
where an adversary repeatedly tries to decrypt this ciphertext \cite{eke}. In order for each party
to ensure the other knows the correct password, they proceed to key verification.

During key verification, the parties exchange messages to prove they each know the session key and,
moreover, the shared password \cite{eke}. In EKE, each party, beginning with the client, encrypts a
random challenge with $R$ and delivers it to the other party, expecting to receive it, re-encrypted,
in return. The structure of the verification messages prevents the receiver from replaying them to
the sender.

\subsection{Attack Flow}
\label{sec:attack-flow}

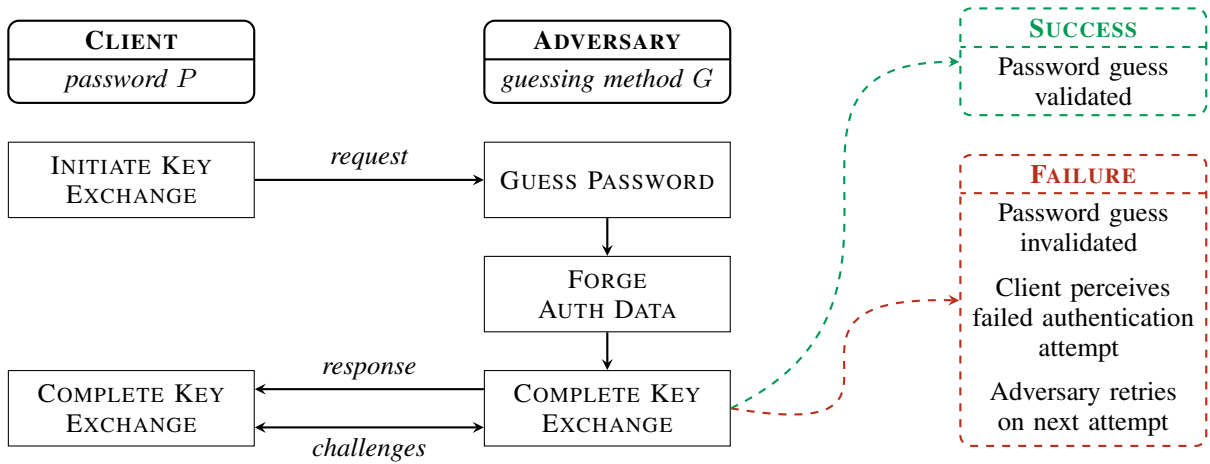
\begin{figure*}[!t]
    \centering
    \begin{tikzpicture}[
    pstep/.append style={font=\scshape},
    result/.style={block, rectangle split, rectangle split parts=2, rounded corners, thick, dashed},
    branch/.style={thick, dashed, -stealth},
    branchlabel/.style={rotate=30, align=center, fill=white}
]
    \node (client) [party] {\textbf{\textsc{Client}} \nodepart{two} \textit{password $P$}};
    \node (adversary) [party, right=of client] {\textbf{\textsc{Adversary}} \nodepart{two} \textit{guessing method} $G$};

    \node (init) [pstep, below=of client] {Initiate Key Exchange};

    \node (guess) [pstep, below=of adversary] {Guess Password};
    \node (forge) [pstep, below=of guess] {Forge Auth Data};
    \node (ashake) [pstep, below=of forge] {Complete Key Exchange};

    \node (cshake) [pstep] at (init |- ashake) {Complete Key Exchange};

    \node (success) [result, draw=ForestGreen, right=of adversary] {
        \textbf{\textsc{\textcolor{ForestGreen}{Success}}}
        \nodepart{two}
        Password guess validated
    };
    \node (failure) [result, draw=BrickRed, below=of success] {
        \textbf{\textsc{\textcolor{BrickRed}{Failure}}}
        \nodepart{two}
        Password guess invalidated \vspace{.2cm} \\
        Client perceives failed authentication attempt \vspace{.2cm} \\
        Adversary retries on next attempt
    };

    \draw[thick, -stealth] (init) -- (guess) node [midway, above] {\textit{request}};
    \draw[thick, -stealth] (guess) -- (forge);
    \draw[thick, -stealth] (forge) -- (ashake);
    
    \draw[thick, -stealth]
    ($(ashake.north west)!0.25!(ashake.south west)$) -- ($(cshake.north east)!0.25!(cshake.south east)$)
    node[midway, above] {\textit{response}};
    
    \draw[thick, stealth-stealth]
    ($(cshake.north east)!0.75!(cshake.south east)$) -- ($(ashake.north west)!0.75!(ashake.south west)$)
    node[midway, below] {\textit{challenges}};

    \draw[branch, color=BrickRed] (ashake.east) .. controls (failure.south west) and (forge.east) .. (failure.west);

    \draw[branch, color=ForestGreen] (ashake.east) .. controls (failure.west) and (adversary.east) .. (success.west);
\end{tikzpicture}
    \caption{
        Overview of a reverse online guessing trial. The adversary impersonates a server to trick
        the client into becoming a password oracle.
    }
    \label{fig:attack}
\end{figure*}

Reverse online guessing works by carrying out a series of \emph{trials}, during each of which the
adversary attempts to validate one password guess for a certain user. \autoref{fig:attack} describes
the flow of a trial. An honest client attempts to initiate key exchange with the server, but is
tricked into communicating with the adversary instead. The adversary proceeds in three steps:

\begin{enumerate}
    \item \textbf{\textsc{Guess Password:}} The adversary produces a guess $P'$ of the client's
    password. This guess can be drawn using a method $G$. The exact method is unimportant; the
    adversary may use simple techniques such as guessing from a dictionary of common passwords or
    guessing based on available user information, or more sophisticated techniques such as Markov
    modeling or machine learning models \cite{unixpw, fast-dictionary, omen, passgan,
    repr-learning}.

    \item \textbf{\textsc{Forge Auth Data:}} The adversary simulates the registration phase locally
    to forge the auth data which an honest server would store. For EKE, nothing needs to be forged,
    since the auth data consists of $U$ and $P$ (here, $P'$), but other protocols such as A-EKE
    require more complicated data structures such as a salt $s$ and a hash of $P'$ and $s$.

    \item \textbf{\textsc{Complete Key Exchange:}} Using the forged auth data, the adversary
    proceeds as an honest server would, sending the client an appropriate response and completing
    key negotiation and verification.
\end{enumerate}

If key verification completes successfully, then the adversary can conclude their guess of the
client's passsword is correct, and the client believes that they have authenticated with the server.
Since the client acts as a password oracle and may have an incorrect password (due to user error),
an adversary should authenticate with the honest server to \emph{truly} verify their guess.
Even a guess rejected by the server is still useful. Assuming it is close to the true password, the
adversary can indicate a ``server-side error'' to the client, urge them to reauthenticate, and use
the bad guess to improve their future guesses, increasing their odds of success%
\footnote{Invalidated credentials may be valid on other servers, so the adversary could also attempt
a \emph{credential stuffing attack} to compromise the same user's accounts elsewhere
\cite{stuffing}}.

On the other hand, if an error occurs during key verification, it may be due to a network failure,
an implementation bug, or an incorrect guess by an adversary (which, to the client, would be
indistinguishable from a misentered password). In interactive login settings, a user would likely
conclude that this was due to an innocuous error and try logging in again. This gives the adversary
another opportunity to guess which they may do as many times as the user tries to log in.

\subsection{Adapting to Other Protocols}
Nothing in the above flow is specific to EKE: reverse online guessing attacks can target a variety
of other PAKE protocols. In fact, in protocols where the auth data is derived from server-generated
values, such as salts, the forged auth data will likely differ from what the honest server has
stored, but the client will be unable to tell the difference.

With a straightforward adaptation of \autoref{fig:attack}, we have been able to demonstrate reverse
online guessing attacks against A-EKE \cite{aeke}, Secure Remote Password (SRP) \cite{srp3,
srp6a-rfc}, OPAQUE \cite{opaque}, Dragonfly Key Exchange \cite{dragonfly}, and Owl \cite{owl}. (As
discussed in \autoref{sec:symbolic-analysis}, these vulnerabilities can be exposed with tools for
symbolic protocol analysis.) Rather than describing each variant in detail, we'll content ourselves
with a brief enumeration of the most interesting issues raised by each protocol.

\subsubsection{Secure Remote Password}
\label{sec:srp}
Like other protocols, SRP-3 \cite{srp3} is vulnerable due to a lack of server authentication. This
issue can be prevented in SRP-6a, as incorporated into TLS 1.2 in RFC 5054, if the client and server
agree to a cipher suite which requires the server to authenticate itself with a certificate and
digital signature \cite{srp6a-rfc}. However, if the client does not require one of these modes, it
remains vulnerable to attack.
Although RFC 5054 has never been deployed at large, and does not apply to TLS 1.3, it remains a
useful case study for employing and preventing reverse guessing attacks.

\subsubsection{OPAQUE}
\label{sec:opaque}
OPAQUE \cite{opaque} is noteworthy for the intricate structure of its auth data, which includes an
asymmetric keypair generated by the client that is further encrypted with a hash of the password.
This hash is collaboratively computed by the client and server with the use of an \emph{oblivious
pseudorandom function} in which the client knows only the password input and hash output while the
server knows only the salt input.
Unfortunately, as existing literature explains (cf. \autoref{sec:related-work}), this complexity
cannot replace a PKI in preventing forgery.

Note that OPAQUE need not just be an PAKE protocol, but can be ``a general secret retrieval
mechanism''; the protocol can be adapted so that the client entrusts the server to store an
arbitrary secret instead of an asymmetric keypair \cite{opaque}. It follows that reverse online
guessing attacks could be adapted to feed falsified data to unsuspecting clients.

\subsubsection{Owl}
Owl \cite{owl} was designed to replace OPAQUE \cite{opaque} and SRP-6a \cite{srp6a-rfc} to avoid
standing issues with the analysis and standardization of these protocols. The authors prove a
\emph{session key indistinguishability} property which guarantees that an adversary who impersonates
an honest server cannot learn the client's session key if they guess the client's password
incorrectly. Moreover, they learn nothing about the password from a leaked session key.
Unfortunately, these properties do not guarantee anything if the attacker \emph{does} guess the
password correctly, perhaps after several attempts, so the protocol remains vulnerable to reverse
online guessing.

\section{Comparison with Standard Online Guessing}
\label{sec:comparison}

Superficially, it might seem that reverse online guessing does not warrant any special treatment
compared to its standard counterpart, as evidenced by the rare references that have acknowledged its
existence (cf. \autoref{sec:related-work}). On the contrary, while these attacks are not inherently
more complex, they present unique challenges for the parties involved and may pose a more serious
threat.

\subsection{Detecting an Attack}
An online guessing attack, reverse or not, requires a significant number of guesses, most of which
will fail, because each trial has a low probability of success. Such a large amount of failed
authentication attempts may indicate an attack. However, the standard attack remains far easier to
detect because a server can track incoming login attempts, especially failing ones, and use this
information to stop a suspected attack. Such mechanisms are an active area of research \cite{pgrp,
StopGuessing, arana}.

Unfortunately, these techniques do not apply to a reverse attack because the server no longer
receives a bulk of authentication requests from the adversary. Instead, the burden of detection is
shifted to the clients.
If the adversary attacks a client and it repeatedly fails to authenticate, the user may grow
frustrated and give up. A vigilant user may grow suspicious, particularly if they expect that their
account should become locked out at some point.
However, clients do not have the same remedies available, particularly automated ones, as a
centralized server does to detect or halt an in-progress attack. Moreover, since the clients are
distributed, each is unaware of what the others, which may also be under attack, are experiencing.
Even if one client detects the attack and protects itself, the adversary can continue to attack
other clients, whereas when a server detects an attack, it can stop it outright and protect every
client or user.

To fly even further under the radar, the adversary can artificially limit the number of guesses per
client to avoid suspicion by identifying clients by their IP address, user agent string, username,
or other information. This limit can be fixed or randomized for each client and reset after a time
period of the adversary's choosing.
In this way, the adversary can distribute a massive amount of guesses across a range of clients
while remaining entirely imperceptible to the server. With only a few trials per client, each user
is unlikely to notice an attack---and the honest server even more so.

As long as the adversary is uninterested in targeting specific clients or users, but rather a large
group indiscriminately, these limits are hardly an inhibitor for them. In an online attack where an
adversary guesses a small number of popular passwords against many accounts, passwords are estimated
to provide about 10 bits of security. This means that with 10 guesses per account, an adversary
would compromise about 1\% of accounts \cite{guessing-science}. Although the exact numbers will vary
depending on the system under attack, how the adversary sources their guesses, and how
security-conscious the user base is, this demonstrates that these attacks can still be used to
compromise many credentials.

\subsection{Bypassing Multifactor Authentication}
\label{sec:mfa}
Once an adversary determines a password, they likely want to impersonate an honest client in order
to gain access to sensitive information or perform actions on their behalf, without their knowledge.
However, some servers may also require \emph{multifactor authentication} (MFA) following password
authentication \cite{nist-dig}. 

With only a password, an adversary will be unable to pass this step without the client's help.
In a standard attack, the adversary will need to launch a separate attack to advance past this
point. Furthermore, they risk detection if the server notifies the user of a new login, perhaps via
an email or push notification.
However, in a reverse attack, the affected user was trying to login anyway and will gladly help the
adversary authenticate.
One-time passwords (OTPs), a common MFA option, afford no additional security here. An adversary
could easily behave as a \emph{man-in-the-middle} (MITM) and forward an OTP between the honest
client and server. Out-of-band push notifications are no help either; in fact the adversary needn't
intervene at all for these.

Moreover, depending on the system under attack, the adversary may be able to request that the server
``remember their device'' so that the adversary's future logins with the same credentials do not
require multifactor authentication and can be done without the user's aid.


\section{Practical Attacks}
\label{sec:practical-attacks}
To illustrate the practical ramifications of reverse online guessing, we enumerate several scenarios
where the attack could be effective.


\subsection{Phishing and Pharming Attacks}
Phishing and pharming attacks are a common way to steal user passwords. An adversary directs an
unsuspecting user to a fraudulent webpage under their control, the user enters their credentials or
other sensitive information into the webpage, and the webpage delivers these in the clear to the
adversary \cite{nist-dig}.

As noted by Hao and Oorschot, ``PAKE protocols appear naturally resistant to phishing attacks''
\cite{pake-sok}. Rather than trusting an authentication interface which the server provides to
handle the password securely (particularly if it runs JavaScript), the client can use a specialized
authentication interface to collect a password and force the server into participating in a PAKE
protocol flow. For example, the client's web browser could include a trusted interface component
that mediates authentication with third-party services via a PAKE protocol. Since PAKE protocols do
not expose passwords by design, this would be enough to rule out standard phishing attacks.

Despite this, an adversary who lures a PAKE client into initiating authentication with them could
use reverse online guessing to bypass some of the client's protection. One possible way is to
corrupt DNS or other infrastructure; for instance, prior work has shown how an adversary can
exploit shared DNS infrastructure to redirect traffic to a spoofed domain \cite{xdauth}. Since most
guesses will fail, this would be less effective than a standard phishing attack but could still lead
to several compromised accounts.

\subsection{Password Spraying Attacks}
\label{sec:spraying}
One popular type of online guessing attack is a \emph{password spraying attack} where an adversary
sources their guesses from a small dictionary of common, weak passwords. Since the dictionary is
small, the adversary only performs a few guesses per user account. These attacks have been used by
several state-level espionage groups and incorporated into different malware packages in the past
\cite{spraying}. Since reverse online guessing may only offer adversaries a few guesses per user, it
is a natural vehicle for conducting a password spraying attack.


\subsection{Compromising Wireless Networks}
\label{sec:attack-wifi}
Reverse online guessing may be especially devastating for wireless network security. The
WPA3-Personal standard employs the Dragonfly PAKE protocol to authenticate client devices
\cite{wpa3}, which unfortunately is vulnerable to reverse online guessing attacks.
In this setting, a reverse guessing attack could facilitate an \emph{evil twin attack} in which an
adversary establishes a network with the same profile (network name and security parameters) as a
trusted network, broadcasts it with a stronger signal, intercepts incoming connection requests, and
tricks client devices into connecting to their network \cite{wpa3, karma}.
A more sophisticated variant, called a \emph{KARMA attack}, determines the ideal network to
impersonate by taking advantage of probe messages which client devices broadcast as they search for
their trusted networks \cite{karma}.
These attacks often occur without user intervention because the fraudulent network matches the
profile of a trusted network which the client will automatically connect to.

Moreover, since WPA3-Personal uses one password to authenticate every station, each authentication
request provides the adversary with a new opportunity to guess the network password. Once they have
deduced it from one device, they can trick other devices into connecting to their network without
further guessing.
Furthermore, wireless networks can have many connected devices, such as desktops, laptops, printers,
smartphones, smart TVs, smart speakers, or Internet of Things devices. With each of these devices
available, all sharing one password, an adversary could accumulate enough guesses to compromise a
weak password.
As opposed to other applications of reverse online guessing attacks where it is necessary to target
users indiscriminately, this can allow an adversary to target a specific wireless network.
On the other hand, this could lead a user to notice something is wrong if many of their devices are
offline, so it may help to time the attack appropriately (such as at night or while users are away
from home during the workday).

A device can protect itself from compromise by using SAE-PK to connect to a network, which enhances
WPA3-Personal by requiring the access point to authenticate itself with a digital signature.
However, if the device does not mandate SAE-PK, but also accepts basic SAE, an adversary can add in
a downgrade attack, leaving the device no better off \cite{wpa3}.

\section{Mitigating the Threat}
\label{sec:mitigation}
Although reverse online guessing poses a venerable threat, there are some ways to mitigate its
effects and prevent it outright.
A direct approach is for clients to mandate that the remote server authenticate itself with digital
signatures. As shown with certain modes of RFC 5054 (cf. \autoref{sec:srp}) and WPA3-Personal (cf.
\autoref{sec:attack-wifi}), this can prevent the attack outright, since forging a digital signature
is much harder than guessing a weak password. Trust in these signatures may rely on PKI or saving
the server's public key to the client's memory during registration.

Whereas signatures are designed to verify the server's identity, it is also possible to verify the
contents of its responses. The OPAQUE protocol \cite{opaque} requires a server to store an encrypted
envelope constructed by the client. To ensure the envelope is authentic, the client could store
secret data inside of it and check that this data is correct during key verification. If the check
fails, the client can proceed like they would in the event of a mistyped password (e.g. dropping the
connection), which would prevent the server from learning that their guess was correct.


To this end, rather than asking the user to remember another password, which would be burdensome,
the protocol could ask users to remember an image or other recognizable information. This concept
was implemented as part of the SiteKey system previously used by some banking websites to help users
notice if they had entered a phishing website.
Although the idea makes some intuitive sense, it had several glaring flaws. A phishing website could
simply omit the picture entirely without the user taking notice \cite{emperor}. Users who
\emph{would} take notice could be tricked too; the phishing website could use a man-in-the-middle
attack to scrape the user's confirmation image from the honest website, as the image was displayed
\emph{before} password entry \cite{sitekey-mitm}. These issues eventually led to SiteKey's
discontinuation.

Nevertheless, the deployment of PAKE presents the opportunity to reevaluate some of these issues. In
particular, if the authentication interface is controlled by the client, as is often the case for
PAKE \cite{pake-sok}, it can force the inclusion of a confirmation image to ensure the user always
sees it, thus restoring some amount of protection. Moreover, it would not be possible to learn the
honest confirmation image from the honest server without already knowing the password or
bruteforcing the server with a traditional online guessing attack, making the whole endeavor
fruitless. Even if some users ignore red flags, vigilant users may appreciate the protection.

\section{Symbolic Analysis}
\label{sec:symbolic-analysis}
Careful protocol analysis can reveal PAKE protocols which are immune from or vulnerable to reverse
online guessing attacks. To simplify this, we propose a technique based in the symbolic model of
cryptography introduced by Dolev and Yao (DY) \cite{dy}.
As a result, this technique applies to abstract protocol specifications devoid of specific
cryptographic primitives which may be found in an implementation. Moreover, the adversary in this
model is implicitly assumed to be capable of manipulating messages sent over public communication
channels, thereby capturing how they might intercept authentication requests from clients.
These details of the abstract model make it easy to encode a protocol, particularly in a software
checker.


Analyzing a protocol of interest begins with encoding it into the symbolic model in a tool of
choice. There are no real restrictions on the encoding, so long as it is a faithful representation.

The next step is to represent how the adversary can guess passwords. One caveat of the symbolic
model is that adversaries are generally unable to guess values, even low entropy ones. Thus, there
is no direct encoding of password guessing. Instead, to check for a viable guessing attack, the
password needs to be leaked during the registration phase, essentially before authentication begins.
Although this may seem to trivialize any subsequent attack, it is a useful way to represent correct
guesses. In particular, this leak can be used to demonstrate protocol flows which an adversary with
a correct password guess can induce.

Finally, check that, in all possible protocol traces, each time the client has completed key
verification of a session key $K$, the server has also done this, with the same key. A protocol
which is weak to reverse online guessing attacks will allow for a client to complete key exchange
with the adversary, entirely without the server's participation.


We have used this technique with the help of two automated protocol verifiers in the symbolic model:
ProVerif\footnote{ProVerif is available at \url{https://proverif.inria.fr}} \cite{proverif} and
CPSA\footnote{CPSA is available at \url{https://github.com/mitre/cpsa}} \cite{cpsa}. These have
diagnosed the flaw in a number of protocols, including EKE \cite{eke}, A-EKE \cite{aeke}, SRP
\cite{srp3, srp6a-rfc}, Dragonfly \cite{dragonfly}, and OPAQUE \cite{opaque}.

\subsection{An Illustration of the Technique}

To demonstrate the technique, we apply CPSA to RFC 5054 \cite{srp6a-rfc}, "Using the Secure Remote
Password (SRP) Protocol for TLS Authentication." We choose to illustrate the technique with RFC 5054
because it supports authentication of the server with and without a certificate. This allows us to
demonstrate the attack and an effective mitigation, server authentication with a certificate, using
our technique. We model RFC 5054 in TLS 1.2 as modeling the RFC in TLS 1.3 would preclude the attack
given that server certificates are always required.

CPSA is different from many other symbolic analysis tools in that it operates as a model finder.
Given a protocol definition and an initial set of assumptions, CPSA will search for all possible
executions of the protocol that satisfy these assumptions. This is particularly useful for
identifying attacks, such as the reverse password guessing attack we have identified. Additionally,
CPSA includes tools that allow the visualization of the theorems it discovers as graphs of protocol
executions, similar to protocol diagrams found in specifications. We will make use of the diagrams
to illustrate the attack. 

RFC 5054 uses SRP-6a. The protocol assumes that the client has registered a verifier, derived from
the client's password, with the server. The verifier is generated as a Diffie-Hellman value
consisting of a generator raised to the value of a hash of a salt, provided by the server, with a
hash of the user's identity and password. The server then stores the verifier and salt with the
client's identity to authenticate the client. TLS is then modified in the following ways to make use
of SRP-6a:
\begin{itemize}
    \item{The client's hello message includes the client's identity.}
    \item{The server includes a server key exchange message that contains the Diffie-Hellman group
    parameters, the client's salt, and a fresh Diffie-Hellman public key chosen by the server. (If
    the server includes the optional server certificate, the key exchange message also includes a
    signature on the hash of the key exchange message.)}
    \item{The client sends a client key exchange message consisting of a fresh Diffie-Hellman public
    key chosen by the client.}
    \item{Both parties then compute a premaster secret based on the SRP key exchange. (See
    \cite{srp6a-rfc})}
\end{itemize}

To model the attack, we model the initial secure exchange of the verifier between the client and the
server (\autoref{fig:init-exchange}). We use CPSA's state mechanisms to store the password in the
client's memory (clientmem) and the verifier in the server's password database (passwdDB). CPSA
cannot use hashes as exponents, so we model the exponent derived from the password as simply a fresh
exponent $x$. CPSA uses the predicate (uniq-gen x) to specify that the value does not exist before
its first appearance in the protocol and therefore cannot be known to the adversary unless the
protocol leaks it. To secure the exchange of messages and prevent leaking of the registration, we
use the ltk function that takes two values of type name and produces a unique symmetric key for the
pair of name values. The first four messages in the initialization exchange create the verifier and
store the password and verifier securely in the client's memory and the server's password database,
respectively. In a usual symbolic analysis, there would be no further messages, but in this case we
add one additional message to the server-init role, (send (exp (gen) x)), releasing the verifier to
the adversary. This last message represents the case where the adversary makes a correct password
guess and can generate a verifier, since no regular participant in the protocol is looking for that
message. This extra message is unique to our technique for detecting the reverse online password
guessing attack.

\begin{figure}[h!]
\small
\begin{minted}{lisp}
  (defrole client-init
    (vars
     (s text) 
     (x rndx) 
     (client server name)
     (clientmem locn)) 
    (trace
     (send (enc client (ltk client server)))
     (recv (enc s (ltk client server)))
     (stor clientmem
           (cat "memory" s x client server)) 
     (send (enc (cat "Enroll" (exp (gen) x))
                (ltk client server)))
     )
    (uniq-gen x)
    )

  (defrole server-init
    (vars
     (s text)
     (x expt) 
     (client server name)
     (passwdDB locn))
    (trace
     (recv (enc client (ltk client server)))
     (send (enc s (ltk client server)))
     (recv (enc (cat "Enroll" (exp (gen) x))
           (ltk client server)))
     (stor passwdDB
           (cat "Client record" client
                s (exp (gen) x) server))
     (send (exp (gen) x)) 
     )
    (uniq-orig s)
    )
\end{minted}
\caption{The initial exchange between a client and a server to establish a secret verifier based on
the client's password at the verifier. It includes a final send in the server-init role that
releases the verifier to the adversary to simulate a correct password guess.}
\label{fig:init-exchange}
\end{figure}

The models for the two versions of TLS 1.2 using SRP, with and without a server certificate, can be
found in the shared artifact. Of particular note are the differences in the server key exchange
messages between the two versions. In the version without a certificate, the server key exchange
contains only the SRP values, but in the version with a certificate, the server sends a certificate
and also includes a signature on the SRP values in the key exchange to authenticate the server to
the client. 

To analyze the models, we provide an initial set of assumptions for the client role and allow CPSA
to determine how many essentially different executions exist that will complete the role. For both
models, we assume that for all clients completing a run of the TLS handshake, the passwords are
uniquely generated during the initialization, a strictly stronger assumption than reality, the salts
are uniquely generated for each client initialization, the long term keys between the clients and
the servers used during an initialization are unknown to the adversary and never transmitted in a
message that would allow the adversary to obtain them, and that the clients' random nonces in the
hello messages are also freshly generated. For the model with a certificate, we also assume that the
servers' and the certificate authorities' private keys are uncompromised.

\begin{figure}[t]
\centering
\def\svgwidth{0.6\columnwidth}
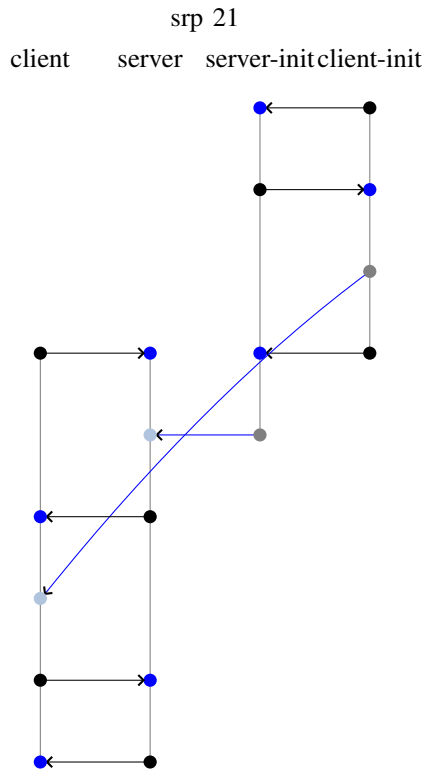
\caption{Run of TLS 1.2 using SRP where the password is not correctly guessed by an adversary. Client completes run with the
server with injective agreement on all variables as indicated by the solid lines between the roles.}
\label{pwgshape1}
\end{figure}

\begin{figure}[t]
\centering
\def\svgwidth{0.6\columnwidth}
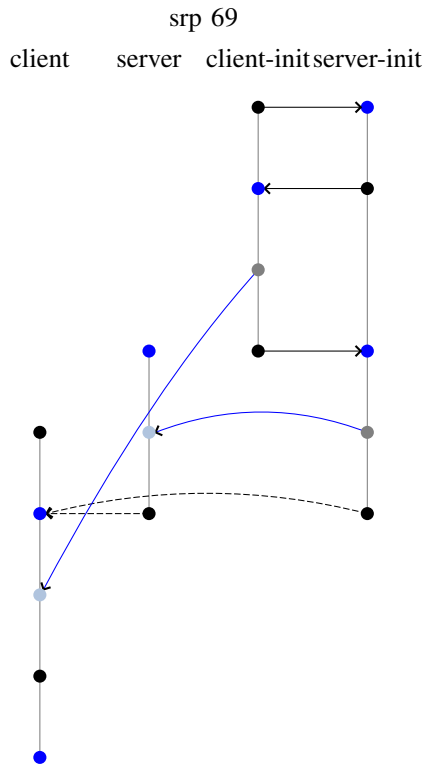
\caption{Run of TLS 1.2 using SRP where the password is correctly guessed by an adversary as shown
with the fifth message from the server-init role. Client completes run without the server indicating
an attack. The client does not know with whom the protocol completes, but believes it completed with
the server.}
\label{pwgshape2}
\end{figure}

The analysis of TLS 1.2 without a server certificate produces multiple possible executions. The
model was not run to termination as it is not necessary to find the attack. The model produced two
distinct classes of executions, ones where the adversary did not correctly guess the password, and
ones where the password was correctly guessed. In those where the password was not guessed correctly
(\autoref{pwgshape1}), the client achieves injective agreement on all variables with a server run of
the protocol. This can be seen by the client connecting to a complete run of the server with solid
lines. The solid lines indicate that the messages that were received were the same messages that
were sent. Both the client and the server agree on the values of all variables in the messages.
Where the password is guessed (\autoref{pwgshape2}), it can be seen that the client completes
without a server run of the protocol, illustrating the attack. Instead of a complete run with the
server, the extra message that makes the verifier available to the adversary added to the
server-init role to simulate a correct password guess appears. In this case, there is a dashed arrow
from the fifth message of the server-init role. This dashed arrow indicates that the adversary can
use the provided verifier to create the message received by the client, allowing the adversary to
impersonate the server.

\begin{figure}[h]
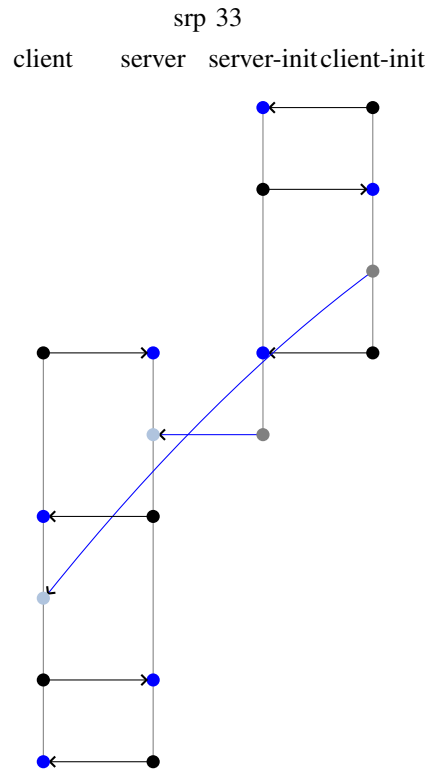

\centering
\def\svgwidth{0.6\columnwidth}

\caption{Run of TLS 1.2 using SRP with server certificate. Client completes run with the server.
CPSA terminated with only this possible execution verifying that server authentication with PKI
mitigates the attack.}
\label{srp6ashape}
\end{figure}

When the same protocol is run with a server certificate, the CPSA analysis completes with only one
class of executions, the class where clients achieve injective agreement with a server
(\autoref{srp6ashape}). As the CPSA analysis terminated, there are no other executions possible for
the protocol. This demonstrates that the addition of server authentication eliminates the reverse
online guessing attack.

The above examples demonstrate our technique for using symbolic protocol analysis tools to search
for the reverse online password guessing attack and to verify mitigations of the attack. The use of
PKI to mitigate the attack is one such possible solution, as proven above. Other possible solutions
could be tried and tested using the above techniques.

\section{Conclusion}
\label{sec:conclusion}
Despite being underestimated in the literature, reverse online guessing attacks pose a venerable
threat to password security. They are conceptually simple, apply to several practical attack
scenarios, and are far more difficult to detect than traditional online guessing attacks. Our
analysis suggests that PAKE standardization efforts should incorporate explicit server
authentication. PKI is one mechanism, but other approaches are possible (cf.
\autoref{sec:mitigation}), especially for deployments which may need to operate PKI-free.

Considering that these attacks are known, it is important to consider why they have been largely
overlooked. The few previous analyses (cf. \autoref{sec:related-work}) treat them more as technical
possibilities but miss that their key power in practice is how difficult they are to detect; while
servers take note of authentication failures and process them to detect attacks, clients do not,
instead assuming that they are caused by user error. Moreover, the adversary can distribute their
attack against so many clients, each unaware of the others, that hardly any will notice. 

In some cases, this attack has been missed entirely for protocols which have undergone rigorous
analysis to verify their security properties. Both OPAQUE \cite{opaque} and Owl \cite{owl} have
Universal Composability (UC) \cite{uc} proofs and SRP-3 \cite{srp3} has CPSA proofs
\cite{srp3-proofs}.
However, the models in these analyses assume that the password is unknown to the adversary. While
this is usually a reasonable assumption since an adversary with the password can trivially
impersonate the client, it inhibits discovery of attacks such as this one.
By contrast, analyzing the protocol with a new model, with an explicit password leak, reveals this
attack in vulnerable protocols and proves its impossibility in secure ones, such as SRP-6a with
server certificates \cite{srp6a-rfc}.
Ultimately, this suggests that more holistic protocol analyses, applying multiple models in tandem,
may be necessary to more thoroughly vet how a protocol may be threatened or misused; even something
which appears trivial can point out useful insights.

\cleardoublepage

\appendices

\section{Proactive Prevention of Harm}

\subsection{Research Process}
In our research process, we rediscovered reverse online guessing attacks while symbolically
analyzing theoretical models of different PAKE protocols. With knowledge of this attack, we
determined its possible impacts by finding current and proposed applications of these protocols and
determining how an attack might be used in each scenario, particularly in combination with existing
attacks used to compromise weak passwords. We did not attempt to deploy this attack in any setting
or create software which would be able to do so.

\subsection{Mitigating Harm}
At the moment, the main potential harm of our work would be to release a publication without
providing vendors the opportunity to deploy mitigations and improve the security of their products.
To reduce this harm as much as possible, we have begun responsible disclosure proceedings so that
vendors who apply affected protocols are aware of the risks involved and can consider incorporating
our proposed mitigations (cf. \autoref{sec:mitigation}). Users of affected services will benefit
once security upgrades are made.

As this vulnerability applies to a wide range of protocols, there is no one vendor to reach out to.
To reach as many vendors as possible, we have reached out to the IETF Security Area Directors about
this issue, who have forwarded it and a draft of our paper to the Crypto Forum Working Group which
is responsible for designing and incorporating several PAKE protocols. They have also looped in the
CERT team at Carnegie Mellon University in order to coordinate a response with affected vendors.

\subsection{Decision}
In light of our analysis, we decided to go forward with completing our research as well as
submitting this paper for publication. The research process itself has had no direct impacts besides
developing our own knowledge as the research team. We plan to share this knowledge via responsible
disclosure followed by publication. As elaborated above, this provides benefits by educating others
about the risks of password-only protocols, hopefully leading vendors to deploy mitigations to their
existing systems and encouraging protocol designers to consider these potential security flaws in
future protocol designs. Moreover, this may encourage others to develop more mitigation strategies
which are applicable to defending against reverse online guessing attacks.

Should we find that more time is necessary to reach all affected vendors and provide them with time
to roll out mitigations as they see fit, we will request the paper be embargoed.

\cleardoublepage

\section{Open Science}
In accordance with the open science policy, we provide a single artifact sufficient to reproduce and
evaluate the analyses reported in this paper. The artifact is a pre-built Docker image named
\texttt{proverif-container}, which encapsulates all of our source code with for symbolic protocol
models and the Proverif tool. Reviewers may pull the anonymized image with the following command:

\begin{minted}{bash}
docker pull ghcr.io/proverif-anontools/proverif-container:latest
\end{minted}

The container is configured so that invoking the image with no additional arguments runs ProVerif on
the included OPAQUE protocol model and prints the verification to standard output. Reviewers may
execute the anonymized image with the following command:
\begin{minted}{bash}
docker run --rm ghcr.io/proverif-anontools/proverif-container:latest
\end{minted}

For exploration and additional inspection, reviewers can start an interactive shell in the image and
examine the scripts, models, and binaries directly and to invoke commands manually. The command
below opens an interactive \texttt{bash} session inside the container:

\begin{minted}{bash}
docker run -it --rm ghcr.io/proverif-anontools/proverif-container:latest /bin/bash
\end{minted}

The above commands are POSIX shell commands (e.g., \texttt{bash}). Windows users may run these using
WSL, Git Bash, or Docker Desktop's CLI.

No proprietary software or sensitive materials are required to reproduce the reported results; all
necessary files are included in the image. The registry namespace is anonymized and contains no
author identifiers. We have removed author names and affiliations from the metadata. This anonymized
artifact is publicly accessible and will remain available for the review period.

\cleardoublepage
\bibliographystyle{plain}
\bibliography{refs}

\end{document}